\documentclass[runningheads,orivec,a4paper]{llncs}
\usepackage[utf8]{inputenc}
\usepackage{graphicx}

\newcommand{\myparagraph}[1]{\subsubsection{#1.}}

\title{Partial supervision for the FeTA challenge 2021}
\author{
    Lucas Fidon\inst{1}\thanks{Corresponding author: Lucas Fidon, lucas.fidon@kcl.ac.uk}
    \and Michael Aertsen\inst{2}
    \and Suprosanna Shit\inst{3}
    \and Philippe Demaerel\inst{2}
    \and S\'ebastien Ourselin\inst{1}
	\and Jan Deprest\inst{4,5}
	\and Tom Vercauteren\inst{1}
}

\authorrunning{Lucas Fidon et al.}

\institute{
School of Biomedical Engineering \& Imaging Sciences, King's College London, UK 
\and
Department of Radiology, University Hospitals Leuven, Belgium
\and
Technical University of Munich, Germany
\and
Institute for Women's Health, University College London, UK
\and
Department of Obstetrics and Gynaecology, University Hospitals Leuven, Belgium
}

\usepackage{hyperref}

\begin{document}

\maketitle


\section*{Objective}
The Fetal Brain Tissue Annotation and Segmentation Challenge (FeTA) aims at comparing algorithms for multi-class automatic segmentation of fetal brain 3D T2 MRI.
Seven tissue types are considered~\cite{payette2021automatic}:
\begin{enumerate}
    \item extra-axial cerebrospinal fluid,
    \item cortical gray matter, 
    \item white matter,
    \item ventricular system,
    \item cerebellum,
    \item deep gray matter,
    \item brainstem.
\end{enumerate}
%
This paper describes our method for our participation in the FeTA challenge 2021 (team name: TRABIT).

The performance of convolutional neural networks for medical image segmentation is thought to correlate positively with the number of training data~\cite{bakas2018identifying}.
The FeTA challenge does not restrict participants to using only the provided training data but also allows for using other publicly available sources.
Yet, open access fetal brain data remains limited.
An advantageous strategy could thus be to expand the training data to cover broader perinatal brain imaging sources.
%
Perinatal brain MRIs, other than the FeTA challenge data, that are currently publicly available, span normal and pathological fetal atlases as well as neonatal scans~\cite{fidon2021atlas,gholipour2017normative,hughes2017dedicated,wu2021age}.
However, perinatal brain MRIs segmented in different datasets typically come with different annotation protocols.
This makes it challenging to combine those datasets to train a deep neural network.

We recently proposed a family of loss functions, the \textit{label-set loss functions}~\cite{fidon2021label}, for partially supervised learning.
Label-set loss functions allow to train deep neural networks with partially segmented images, i.e. segmentations in which some classes may be grouped into super-classes.
We propose to use label-set loss functions~\cite{fidon2021label} to improve the segmentation performance of a state-of-the-art deep learning pipeline for multi-class fetal brain segmentation by merging several publicly available datasets.
To promote generalisability, our approach does not introduce any additional hyper-parameters tuning.

\section*{Methods and Materials}

\begin{figure}[t]
    \centering
    \includegraphics[width=\textwidth,trim=0cm 2cm 0cm 0cm, clip]{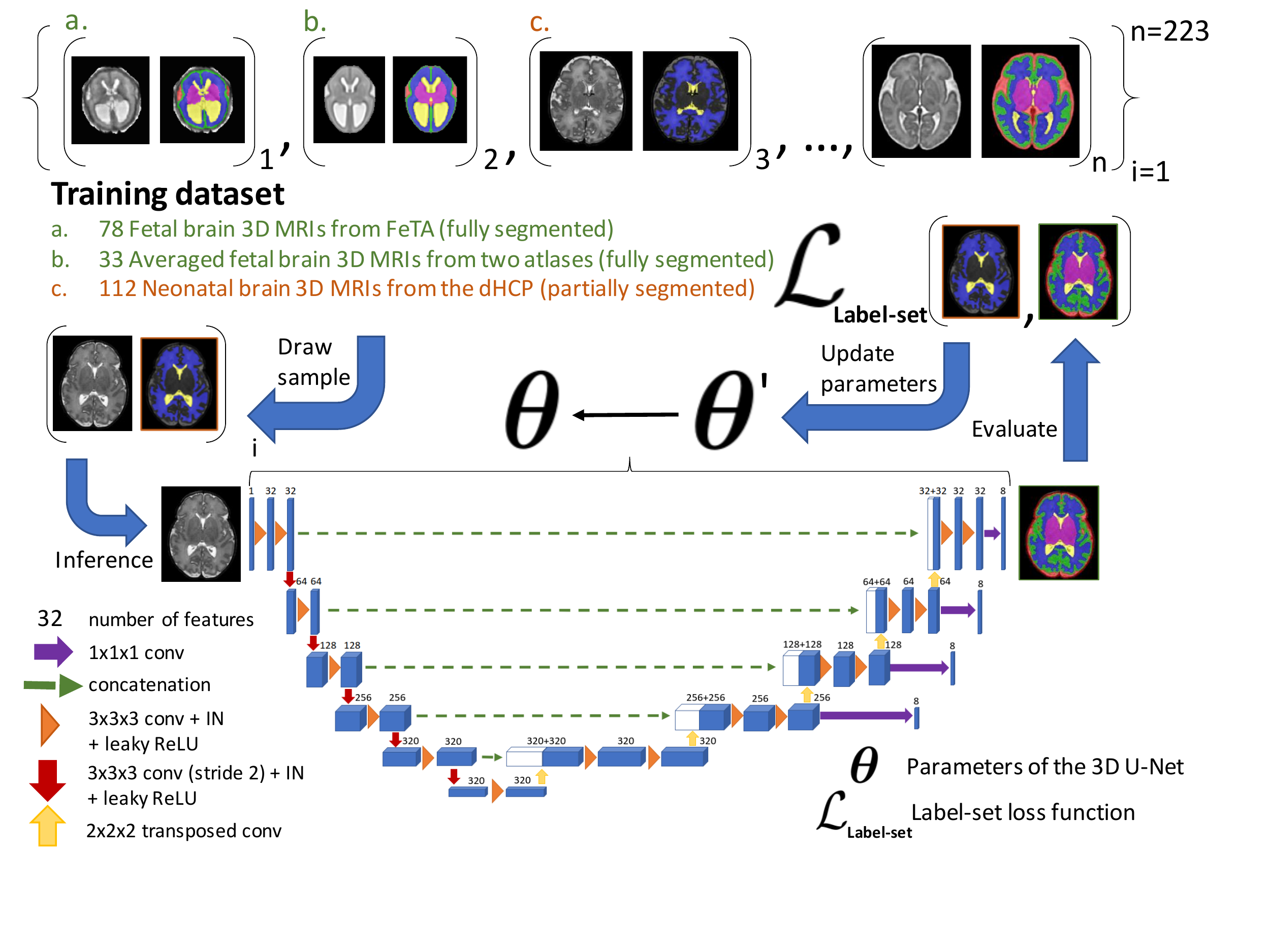}
    \caption{
    \textbf{Overview of the training process with partial supervision.}
    The 3D MRIs from datasets a.~\cite{payette2021automatic} and b.\cite{fidon2021atlas,gholipour2017normative} are fully-segmented while the 3D MRIs from dataset c.~\cite{hughes2017dedicated} were only segmented manually for white matter, ventricular system, cerebellum, and the full brain.
    We propose to use a label-set loss function~\cite{fidon2021label} to allow to train a 3D U-Net~\cite{cciccek20163d} with fully and partially segmented 3D MRIs.
    }
    \label{fig:overview}
\end{figure}

In this section, we give the detail of our segmentation pipeline and the data used for training the deep neural networks.
Our segmentation  software is publicly available at \url{https://github.com/LucasFidon/feta-inference}.

\myparagraph{FeTA challenge amended training data} 
The original FeTA challenge training data provides
$80$ fetal brain 3D T2 MRIs with manual manual segmentations of all
7 target
tissue types~\cite{payette2021automatic}.
$40$ fetal brain 3D MRIs were reconstructed using MIAL~\cite{tourbier2015efficient} and $40$ fetal brain 3D MRIs were reconstructed using Simple IRTK~\cite{kuklisova2012reconstruction}.

For the $40$ MIAL 3D MRIs, corrections of the segmentations were performed by authors MA, LF, and PD using ITK-SNAP~\cite{yushkevich2016itk} to reduce the variability against the published segmentation guidelines that was released with the FeTA dataset~\cite{payette2021automatic}.
Those corrections were performed as part of our previous work~\cite{fidon2021label} and are publicly available (DOI: 10.5281/zenodo.5148612).
Two spina bifida cases were excluded (\texttt{sub-feta007} and \texttt{sub-feta009}) because
we considered that
the image quality did not allow to segment them reliably for all the tissue types.
Only the remaining $78$ 3D MRIs were used for training.

\myparagraph{Other public training data}
We also included $18$ average fetal brain 3D T2 MRIs from a neurotypical fetal brain atlas\footnote{\url{http://crl.med.harvard.edu/research/fetal_brain_atlas/}}~\cite{gholipour2017normative},
$15$ average fetal brain 3D T2 MRIs from a spina bifida fetal brain atlas\footnote{\url{https://www.synapse.org/\#!Synapse:syn25887675/wiki/611424}}~\cite{fidon2021atlas}.
Segmentations for all 7 tissue types are available for all the atlas data.
In addition, we used $112$ neonatal brain MRIs from the developing human connectome project~\cite{hughes2017dedicated} (dHCP data release 2).
We excluded the brain MRIs of babies with a gestational age higher than $38$ weeks.
We started from the brain masks and the automatic segmentations publicly available for the neonatal brain MRIs for white matter, ventricular system, and cerebellum~\cite{makropoulos2018developing} and we modified them manually to match the annotation protocol of the FeTA dataset~\cite{payette2021automatic} using ITK-SNAP~\cite{yushkevich2016itk}.
%
Ground-truth
segmentations for the other tissue types
in the dHCP data
were not available for our training.

\myparagraph{Pre-processing}
A brain mask for the fetal brain 3D MRI is computed by affine registration of template volumes from two fetal brain atlases~\cite{fidon2021atlas,gholipour2017normative}. We use all the template volumes with a gestational age that does not differ to the gestation age of fetal brain by more than $1.5$ weeks.
The affine registrations are computed using a symmetric block-matching approach~\cite{modat2014global} as implemented in \texttt{NiftyReg}~\cite{modat2010fast}. The affine transformations are initialized by a translation that aligns the centre of gravity of the non zero intensity regions of the two volumes.
The brain mask is obtained by averaging the warped brain mask and thresholding at $0.5$.

After a brain mask has been computed, the fetal brain 3D MRI is registered rigidly to a neurotycal fetal brain atlas~\cite{gholipour2017normative} and the 3D MRI resampled to a resolution of $0.8$ mm isotropic.
The rigid registration is computed using \texttt{NiftyReg}~\cite{modat2014global,modat2010fast} and the transformation is initialized by a translation that aligns the centre of gravity of the brain masks.

\myparagraph{Deep learning pipeline}
We used an ensemble of $10$ 3D U-Nets~\cite{cciccek20163d}. 
We used the DynU-Net of MONAI~\cite{monai} to implement a 3D U-Net with one input block, $4$ down-sampling blocks, one bottleneck block, $5$ upsampling blocks, $32$ features in the first level, instance normalization~\cite{ulyanov2016instance}, and leaky-ReLU with slope $0.01$.
An illustration of the architecture is provided in Fig.~\ref{fig:overview}.
The CNN used has $31 195 784$ trainable parameters.
The patch size was set to $128 \times 160 \times 128$.
All the pre-processed 3D MRIs on which the pipeline was tested fitted inside a patch of size $128 \times 160 \times 128$.
Every input volume is skull stripped after dilating the brain mask by $5$ voxels and cropped or padded with zeros to fit the patch size. The non zeros image intensity values are clipped for the values above percentile $99.9$, and normalized to zeros mean and unit variance.
Test-time augmentation~\cite{wang2019aleatoric} with all the combinations of flipping along the three spatial dimensions is performed ($8$ predictions).
The $8$ score map predictions are averaged to obtain the output of each CNN.
Ensembling is obtained by averaging the softmax predictions of the $10$ CNNs.
The deep learning pipeline was implemented using MONAI v$5.2.0$ by authors LF and SS.

\myparagraph{Loss function}
We used the sum of two label-set loss functions as loss function: the Leaf-Dice loss~\cite{fidon2021label} and the marginalized cross entropy loss~\cite{fidon2021label}.

\myparagraph{Optimization}
For each network in the ensemble, the training dataset was split into $90\%$ training and $10\%$ validation at random.
The random initialization of the 3D U-Net weights was performed using He initialization~\cite{he2015delving}.
We used SGD with Nesterov momentum, batch size $2$, weight decay $3\times 10^{-5}$, initial learning rate $0.01$, and polynomial learning rate decay with power $0.9$ for a total of $2200$ epochs.
The CNN parameters used at inference corresponds to the last epoch.
We used deep supervision with $4$ levels during training.
Training each 3D U-Net required $12$GB of GPU memory and took on average $3$ days.
We have trained exactly $10$ CNNs and used all of them for the ensemble submitted to the challenge.
%

\myparagraph{Data augmentation}
We used random zoom (zoom ratio range $[0.7, 1.5]$ drawn uniformly at random; probability of augmentation $0.3$),
random rotation (rotation angle range $[-15^{\circ}, 15^{\circ}]$ for all dimensions drawn uniformly at random; probability of augmentation $0.3$),
random additive Gaussian noise (mean $0$, standard deviation $0.1$; probability of augmentation $0.3$),
random Gaussian spatial smoothing (standard deviation range $[0.5, 1.5]$ in voxels for all dimensions drawn uniformly at random; probability of augmentation $0.2$),
random gamma augmentation (gamma range $[0.7, 1.5]$ drawn uniformly at random; probability of augmentation $0.3$),
and
random flip along all dimension (probability of augmentation $0.5$ for each dimension).

\myparagraph{Post-processing}
The mean softmax prediction is resampled to the original 3D MRI using the inverse of the rigid transformation computed in the pre-processing step to register the 3D MRI to the template space.
This image registration is computed using \texttt{NiftyReg}~\cite{modat2010fast} with an interpolation order equal to $1$.
After resampling, the final multi-class segmentation prediction is obtained by taking the argmax of the mean softmax.

\section*{Results}
We evaluated our method on 20 spina bifida MRIs acquired at University Hospital Leuven that were previously used in~\cite{fidon2021label,fidon2021distributionally} using the Dice score~\cite{dice1945measures,fidon2017generalised}.
Those 20 3D MRIs and their brain masks were computed using the super-resolution and reconstruction software \texttt{NiftyMIC}~\cite{ebner2020automated,ranzini2021monaifbs}.
Our method achieves mean (standard deviation) Dice scores: white matter $92.8$ ($1.8$), ventricular system $93.5$ ($2.9$), cerebellum $86.9$ ($8.5$), extra-axial CSF $81.9$ ($11.2$), cortical grey matter $77.7$ ($8.2$), deep grey matter $81.2$ ($4.8$), and brainstem $74.4$ ($5.0$).


\section*{Conclusion}
Partially supervised learning can be used to train deep neural networks using multiple publicly available perinatal brain 3D MRI datasets that have different level of segmentations available.
We used label-set loss functions~\cite{fidon2021label} to train an ensemble of 3D U-Nets using four publicly available datasets~\cite{fidon2021atlas,gholipour2017normative,hughes2017dedicated,payette2021automatic}.
We have submitted our segmentation algorithm to the FeTA challenge 2021.

\section*{Funding Sources}
This project has received funding from the European Union's Horizon 2020 research and innovation program under the Marie Sk{\l}odowska-Curie grant agreement TRABIT No 765148.
This work was supported by core and project funding from the
Wellcome [203148/Z/16/Z; 203145Z/16/Z; WT101957], and EPSRC [NS/A000049/1; NS/A000050/1; NS/A000027/1].
TV is supported by a Medtronic / RAEng Research Chair [RCSRF1819\textbackslash7\textbackslash34].

%
%
%
\bibliographystyle{splncs04.bst}
\bibliography{main.bib}

\end{document}